# Continuous Spin Fractionation:
# A large scale method to improve the performance of polymers

J. Eckelt[1,2] and B. A. Wolf[1,*]

[1] Institut für Physikalische Chemie der Johannes Gutenberg-Universität,
Welder-Weg 15 D-55099 Mainz, Germany

[2] WEE-Solve GmbH, Auf der Burg 6, 55130 Mainz, Germany

Bernhard.Wolf@Uni-Mainz.de

**ABSTRACT**

Most technical polymers and many biopolymers contain very different molecular species (unlike chain length, molecular architecture and/or chemical composition) in contrast to pure low molecular weight compounds. This inconsistent constitution of macromolecules proves very adverse in many cases. An adequate fractionation of polydisperse polymers is therefore mandatory.

Very efficient means are available for analytical purposes. However, these methods break down as soon as the required amount of product exceeds some ten grams. In order to gain access to large enough quantities of sufficiently uniform polymer samples, we have developed a special kind of extraction process called Continuous Spin Fractionation (CSF). The better soluble macromolecular species are preferentially transferred from a feed phase (concentrated polymer solution) into a pickup phase (solvent of tailored thermodynamic quality). The main problem of the procedure lies in the high viscosities of reasonably concentrated polymer solutions, impeding the attainment of thermodynamic equilibria. This hurdle could be cleared by means of spinning nozzles through which the feed is pressed into the pickup phase. CSF can be implemented to any soluble polymer and is likewise apt for the production of small and of large amounts of polymer samples with the required uniformity. This contribution explains how to customize CSF to the polymer of interest and presents a number of typical examples.

## 1. INTRODUCTION

For polymers of high duty applications it becomes increasingly important to remove undesired components that are unavoidably formed as byproducts in the case of many industrial and biological macromolecules. Because the demand of these specialty polymers does normally by far exceed the scale that can be met by already available analytic fractionation methods, new ways are required to produce the desired amounts. Out of the many separation techniques that are customary for the removal of harmful components from low molecular weight mixtures, only liquid/liquid and liquid/solid phase equilibria can be employed to polymers, where the latter is confined to crystallizable macromolecules. The basis of fractionation in terms of a partitioning of the components on two coexisting phases is well known since the beginning of polymer science as will be shown in more detail in the next section. The problem that needs to be solved in the present context consists in the development of a method that allows the large scale purification of polymers. The first technique that suffices for that purpose was the Continuous Polymer Fractionation (CPF)[29], which consists of a special kind of extraction and can be applied to a multitude of different polymers. However, it suffers from the drawback of working at comparatively high dilution, i.e. requires large amounts of solvents, and is in some cases prone to operating troubles due to the high viscosity of one of the coexisting phases. For that reason we have recently developed a considerably improved method, namely Continuous Spin Fractionation (CSF)[4].

The theoretical basis of the two techniques and essential features of the large scale fractionation methods are presented in the next section. The subsequent part gives some typical examples, demonstrating how synthetic or natural polymers can be fractionated successfully. This contribution concludes with instructions concerning the criteria that need to be considered for the employment of CSF to new types of polymers, which have so far not been dealt with.

## 2. METHODS AND THEIR THEORETICAL BACKGROUND

Because analytical methods must be ruled out for the production of industrially interesting amounts of polymer that do no longer contain harmful components, the only remaining procedures are phase equilibria between condensed phases. Although it is in principal possible to use liquid/solid equilibria for fractionation purposes these techniques are not general due to the fact that most polymers do not crystallize. This means that all large scale fractionation techniques make use of liquid/liquid equilibria. The fundamentals of the distribution of different species of high molecular samples between two coexisting phases have already been studied at the beginning of polymer science [8;26]. Over the years many polymers have been fractionated for basic research, typically up to the order of 10 to 100 g samples; the results of these investigations have been collected in several book dedicated to polymer fractionation[2;9].

For the polymer fractionation on an industrial scale the typical extraction processes employed for the purification of low molecular weight mixtures (cf. Fig. 1) must be ruled out because of the insufficient differences in the solubility of macromolecules with either unlike chain length, architecture or – in the case of copolymers – chemical composition. In Fig. 1 the source phase is called feed (FD) and the extracting agent, thermodynamically more favorably interacting with the polymer is called solvent 2. The two liquids that coexist for a given over-all composition of the mixture (working point: WP) are designated L1 and L2. As shown in this graph all high molecular weight compo-nents are typically found in one of the coexisting phases only, which means that no fractionation takes place.

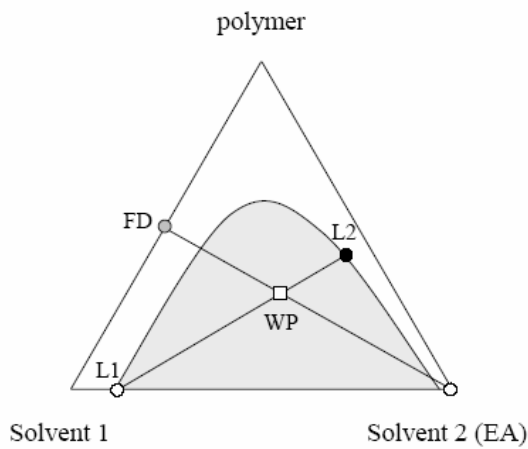

*Fig. 1: Typical phase diagram as used in a conventional liquid-liquid extraction*

For the separation of polymer homologues the query to the components of the mixture must be considerable more subtle. Normally the required two-phase situation is no longer produced by two "immiscible" liquids but by a miscibility gap between the polymer and a single or mixed low molecular weight solvent, as shown in Fig. 2. Fractionation based on solubility is known from the very beginning of polymer science; this separation rests on the fact that an individual distribution coefficient is established for each different species of a non-uniform polymer sample. For macromolecules that differ only in chain length these coefficients are not constant because the Gibbs energy of mixing consists of enthalpic and of entropic parts. The mixture forms two phases: A polymer rich gel phase (GL) and the coexisting polymer lean sol phase (SL). Shorter chains prefer the dilute phase over the concentrated, because of the larger entropy of mixing they can gain. On the other hand, long chains favor the concentrated phase because of the lower number of the energetically very unfavorable contacts to solvent molecules. Although the qualitative features are very clear, a quantitative description of fractionation (i.e. the prediction of the molecular weight distributions of the polymers that are contained in the coexisting phases) is very demanding. The presence of usually thousands of different species and the dependence of the Flory-Huggins interaction parameters on composition and molar mass make the task difficult.

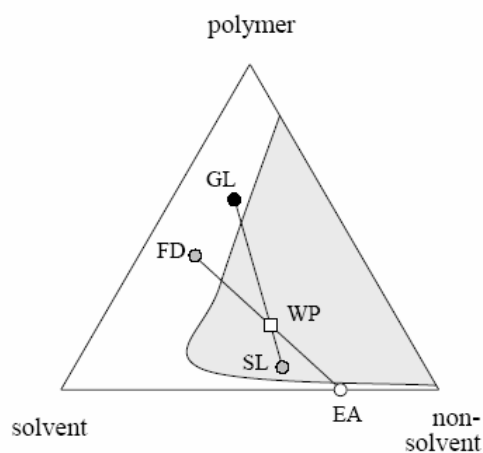

*Fig. 2: Typical phase diagram as used for the fractionation of polymers (reprinted from [5]).*

If the two phase situation is realized with a single solvent, the thermodynamic quality can only be modified by varying the temperature. Although this option can be used for large scale fractionation it is rather inconvenient because the amount of the original polymer that is collected in sol and gel, respectively, is determined by T. For that reason one normally uses mixed solvents and works at ambient temperatures. The relative amounts of the original polymer that will be found in the dilute and in the concentrated phase, respectively, can in this case be tailored by the solvent/non-solvent ratio.

So far we have only dealt with thermodynamic background of fractionation. For practical purposes, in particular in the context of large scale methods it is, however, very important to account for kinetic effects. This becomes immediately obvious if one keeps in mind that the polymer components that need to be removed are initially entangled with all other species and need to be extracted from this source phase into another phase. Even in the case of single solvents, where the demixed state can be reached form the homogeneous mixture by a very slow change in temperature, polymers belonging thermodynamically into the extract are caught in the source phase.

The first method that allowed the production of virtually any amount of polymer fractions was CPF; it consists in a continuous counter current extraction and is normally operated with filled columns. In order to increase the efficiency of fractionation it uses two zones of different temperature, i.e. different solvent quality. The above mentioned difficulties in the establishment of equilibria caused by slow transport processes are at least partially compensated by extracting a larger amount of polymer (i.e. also chains which should be accumulated in the gel phase) at one temperature and precipitating these components at another temperature where the solvent power of the mixed solvent is less. CPF works well as long as the viscosities of the feed phase and of the gel phase remain sufficiently low. If they exceed a critical value there is the danger that the column becomes blocked because of the damming back of these phases.

Because of the described deficiencies of CPF we have developed the advanced technique of CSF. It uses spinning nozzles as employed for fiber production. The decisive advantage of this feature consists in the subdivision of the source phase into many minute particles as shown schematically in Fig. 3. In this manner the distance over which the easier soluble components need to be transported over the phase boundary to reach the continuous phase is minimized and the attainment of equilibria is considerably promoted. How the typical set-up of a CSF apparatus looks like can be seen from Fig. 4. The feed solution containing the polymer to be fractionated is pressed through the spinning nozzles into a vigorously stirred extracting agent (usually a mixture of a thermodynamically good and a bad solvent) and transported in a settler, where sol and gel created by this mixing can separate macroscopically. The location of the composition of the entire liquid contained in the apparatus (the so called working point) within the Gibbs phase triangle (cf. Fig. 2) is at constant temperature fixed by the solvent/non-solvent ratio of the extracting agent and by the ratio of the fluxes of the two entering phases.

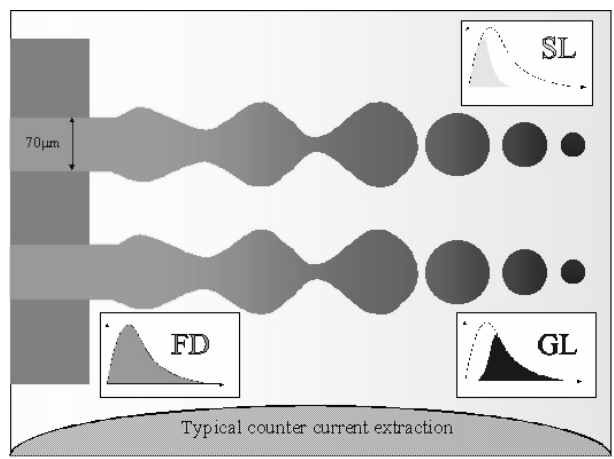

*Fig. 3: Schematic illustration of the breakup of the source phase into minute droplets (reprinted from [5]).*

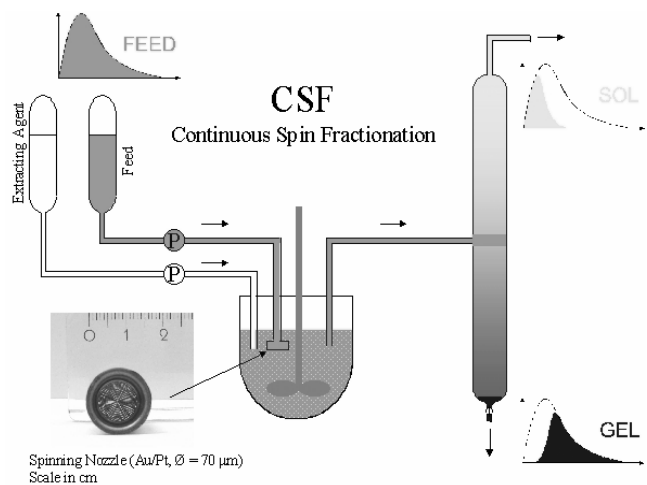

**Fig. 4:** *Typical Set-up of the CSF-Apparatus (reprinted from [5]).*

## 3. EXAMPLES FOR LARGE SCALE FRACTIONATION

### 3.1. SYNTHETIC POLYMERS

Polyisobutylene[10;12] was one of the first systems for which the suitability of CPF for the fractionation of synthetic polymers was demonstrated. In this case we have studied the role of the choice of the best combination of solvent and precipitant systematically. Out of 21 mixed solvents toluene/methyl ethyl keton proved to be best. The fractionation of polyethylene[13] is an example for the possibility to run a large scale fractionation with a single (theta) solvent. In order to establish the required liquid/liquid phase separation one has to work above the crystallization temperature of the polymer and find a system with a suitable Theta-temperature. In the case of polyethylene we used diphenyl ether and worked at 133 and 136 °C respectively. One of the big problems with this polymer consists in the maintenance of the high temperature throughout the entire apparatus, including the pumping device.

Another "well-behaved" homopolymer that was fractionated by means of CPF for scientific purposes was poly vinyl methyl ether[21]. In this case we have used toluene as solvent and petroleum ether as precipitant. Novolak, a photoresist, represents a polymer for which the removal of low molecular weight components is of technical interest in the context of the miniaturization of electrical circuits. How the oligomers can be removed on a technical scale is shown in Fig. 5

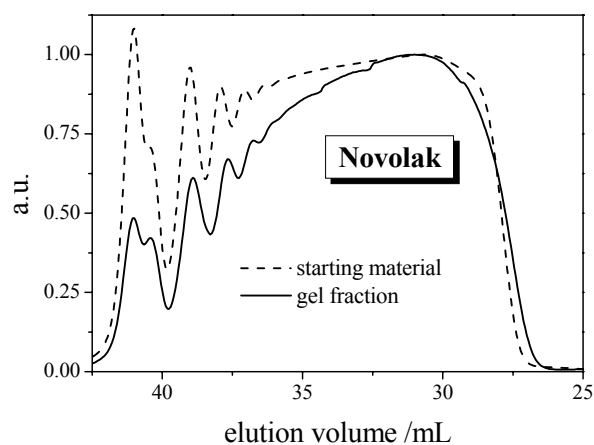

**Fig 5:** *GPC-plot of a Novolak resin before and after the fractionation by mans of CSF (reprinted from [5]).*

Commercial polyacrylic acid normally exhibits a multimodal molecular weight distribution. We have fractionated a sample with three different peaks by means of CPF using the exothermal theta solvent 1,4 dioxane (for which the required two phase state of the solutions is realized at temperatures higher than the theta temperature) plus a filled column and choosing two different temperature regimes for the continuous countercurrent extraction[15]. In this manner it was possible to obtain in two steps a fraction with unimodal distribution and reasonable molecular non-uniformity. Indications exist that the separation does not only take place with respect to chain lengths but also according to the tacticity of the polymer.

Polycarbonate[28] with a molecular non-uniformity $U = (M_w / M_n) - 1 = 1.3$ was the polymer, which served for theoretical considerations concerning the questions, which measures could be taken to improve the efficiency of a simple counter current polymer fractionation. The central outcome of this work consisted in the introduction of the above mentioned temperature zones with different solvent quality. The idea was to extract more than the required amount of easier soluble components at one temperature (making sure that practically no low molecular weight supposed to be contained in the sol fraction is not released from the feed and thus captured in the gel fraction) and to precipitate the undesired higher molecular weight components at the second temperature. By realizing a suitable fractionation strategy and using the gel phase of the different CPF runs as feed for the next extraction it was possible to obtain five fractions of approximately equal weight with $U \approx 0.1$. Melt polymerized bisphenol-A polycarbonate with a low degree of branching was also fractionated[16] by means of CPF; this study has shown that the branching density increases with molecular weight.

All examples of synthetic polymers, except for the last one, referred to strictly linear macromolecules. Recently highly branched products became of great interest. For that reason we are presently separating inevitably formed linear byproducts from the branched material to enable a systematic investigation of the differences in the thermodynamic behavior of linear and branched polymers. In an ongoing thesis[23] we have so far dealt with polyisoprene; the most outstanding observation we made in this context is the fact that linear and branched (typical degree of branching 0.5-0.66) polyisoprene exhibit a large miscibility gap at room temperature.

The large scale fractionation methods discussed here are not confined to homopolymers, as we have for instance demonstrated for random copolymers of styrene and acrylonitrile (SAN). In this case the phase separation can for instance be induced by a suitably chosen second, chemically different polymer such that the separation with respect to the composition of the copolymer becomes dominant as compared with the fractionation according to chain length[20]. In the present case we have used the quasi-ternary system DMAc/SAN/polystyrene, where the solvent dimethylacetamide is completely miscible with both polymers.

Other examples[17] for the suitability of the present methods to fractionate copolymers concern polymer brushes with poly(methyl methacrylate) (PMMA) backbone and polystyrene side chains and hyperbranched PMMA The combination of methyl ethyl ketone (solvent) with acetone (AC = precipitant) turned out to be suitable for the fractionation of the polymer brushes; in case of the hyperbranched material AC served as the solvent component and methanol as the precipitant. A more recent instance of CSF concerns the fractionation of styrene–butadiene block copolymers of different molecular architecture[30]. In this case THF served as the solvent component and ethanol as precipitant.

Polyelectrolytes represent another class of technically important macromolecules for which fractionation is of great interest. CPF experiments were carried out with poly[(dimethylimino) decamethylene bromide] using ethylene glycol monoethyl ether as solvent and diisopropylether as - component[25]. Also fractionated were a cycloaliphatic ionene[24] and an aliphatic random copolymer ionene[22].

Polyvinylchloride[11] turned out to be a polymer that is particularly difficult to fractionate, because of the high tendency of its solutions to gel under the required thermodynamic conditions (finally leading to phase separation). This problem could, however, be solved by deviating from the general rule according to which the differences in the qualities of solvent and non-solvent should be as low as possible. In the case of PVC we could suppress gelation by combining the very good solvent THF with the very strong precipitant water. The reason for this phenomenon lies in the strong preferential solvation of the polymer by THF, which reduces the direct contacts between the individual polymer segments to such an extent that the mixtures remain fluid. From a PVC sample with $M_w$ = 67 kg/mol and U=0.95 we have in four CPF steps obtained five fractions with U ≈ 0.2 and $M_w$ values ranging from 20 to 100 kg/mol.

## 3.2. BIOPOLYMERS AND THEIR DERIVATIVES

The necessity to remove adverse components is especially important for certain types of biomacromolecules, which are by nature synthesized with high molecular or/and chemical non-uniformity. This statement is particularly true for most of the polysaccharides, for which some large scale fractionation experiments are reported below.

Pullulan (450 g, using water as solvent and acetone as precipitant) and dextran (70 g, water plus methanol) with a broad molecular weight distribution were fractioned by means of CSF to demonstrate the suitability of the method for the fractionation of biomacromolecules[7]. Another examples for which the removal of too low and too high molecular weight components is of practical importance is hydroxyethyl starch (HES)[14], which is widely used as blood plasma expander. The short chains are inefficient because their residence time in the body is too short; for the longest chains on the other hand it is too long and the storage of this material in the body causes itching. In this case we used water as solvent and acetone as precipitant.

Cellulose and its derivatives represent another interesting class of polymers for which the large scale fractionation promises important technical progress. From the fact that it requires special solvents and conditions to dissolve cellulose on a molecular scale it is obvious that the task is much easier for the often readily soluble derivatives than for the unsubstituted material. For basic research and in view of the technically important membranes produced form cellulose acetate we have fractionated this material using methyl acetate as solvent component and 2-propanol as precipitant[18]. Using the higher molecular weight fraction of cellulose acetate for the production of membranes it could be shown that it is possible to avoid the formation of "filter dust"[3]. This troubling phenomenon, resulting from short chains that are present in the unfractionated material requires the mechanical removal of the small spherical polymer particles from the surface of the membrane.

In order to prepare cellulose samples with narrower molecular weight distribution we have chosen several strategies. In a first attempt we have prepared trimethylsilylcellulose and fractionated this material[27] in toluene (solvent) and dimethyl sulfoxide (precipitant). Unsubstituted fractions of cellulose can be obtained by desilylation. After that we have checked whether N,N-dimethylacetamide + LiCl or Ni-tren as solvents for cellulose in combination with suitable precipitants can be used[19]. Only the first solvent in combination with acetone proved practical; in Ni-tren the polymer degrades too rapidly[6].

Hyaluronic acid and its salts are of great importance for eye surgeries and in the field of medical care. The molecular weight distribution of this polymer with uncommonly large degrees of polymerization is particularly important for the rheological properties of its aqueous solutions. For this reason and because of the fact that the short chains are suspected to cause inflammations when present in the viscoelastica used in the field of cataract surgeries, their removal appears mandatory. For the fractionation of the sodium salt we have chosen water as the solvent component and 2-propanol as the precipitant. In this case we had to modify the usual CSF procedure because it turned out that the shear fields caused by pressing the feed through the spinning nozzles are high enough to result in the mechanical degradation of this extremely high molecular weight material. The problem could be solved by pressing the mixed solvent into the feed (inverse spin fractionation)[1] instead of the usual practice.

## 4. EMPLOYMENT OF CSF TO NEW POLYMERS

For fractionation the polymer must be soluble and it must be possible to realize liquid/liquid phase separation. In most cases there exists a multitude of low molecular weight liquids that can serve as solvents or as precipitant, i.e. as components of the mixed solvent causing demixing, even at the experimentally most convenient ambient temperatures. Under these conditions the most important preparatory step consists in a systematic investigation of the different combinations of solvents and precipitants. Normally it is advantageous for the separation to combine the thermodynamically least favorable solvent with the weakest precipitant; however, in order to circumvent gelation for polymers of high self association or crystallization tendency it is sometimes necessary to diverge from that rule.

Normally it is possible to find several mixed solvents that are suitable with respect to the thermodynamic requirements. If this is the case the next step consists in the determination of phase diagrams, in particular it is recommendable to measure the densities of the coexisting phases and the critical polymer concentration. For the kinetics of macroscopic separation of the coexisting phases resulting in CPF or CSF it is of course advantageous to use the mixed solvents that result in the largest differences between the densities of the coexisting phases; in case no suitable solvents can be found one employing a centrifuge does normally solve the problem. The value of the critical polymer concentration determines how much of low molecular weight components is required to fractionate a certain polymer sample. The reason is that separation efficiency worsens as the distance between the working point in the Gibbs phase triangle and the critical point of the system shrinks. Another important aspect that helps the optimization of large scale fractionation consists in the viscosities of feed and to a lesser extent of the gel phase. High viscosities reduce the rate of the transfer of matter over the phase boundary but they are also unfa-vorable for the pumping through the spinning nozzles. In case all the favorable items described above are fulfilled by several mixed solvents the one should be chosen for which it is possible to realize a working point such that the angle between tie line and the working line (connecting the compositions of feed and extracting agent, cf. Figs. 1 and 2) results as small as possible.

After all this preparatory work has been successfully accomplished it is essential to develop a fractionation strategy that is adequate to reach the goal, which can for instance consist in the removal of the lowest or/and the highest molecular weight material of a given sample. Where the cut through the molecular weight distribution will be performed can be controlled by the composition of the mixed solvent in feed and in the extraction agent, respectively, and if these data are fixed, by the rate with which these phases are pumped into the apparatus. For the realization of the requested absence of harmful components that cannot be achieved in one step it is very advantageous that there is normally no need to separate the polymer from the gel phase for the next fractionation step. The gel phase can either be directly used as feed for the next CSF or CPF run or it requires only the addition of some solvent or precipitant.

**ACKNOWLEDGEMENT**

We would like to thank the Lenzing AG, Austria for providing us with spinning nozzles.